\documentclass[prd,nofootinbib]{revtex4-2}
\usepackage{amssymb, amsfonts, graphicx, amsmath,color} 
\usepackage{hyperref}

\newcommand{\ket}[1]{|#1\rangle}
\newcommand{\bra}[1]{\langle#1|}
\newcommand{\braket}[2]{\langle #1| #2\rangle}

\newcommand{\SO}{\mathrm{SO}}
\newcommand{\SU}{\mathrm{SU}}
\newcommand{\FP}{{\mathfrak{P}}}
\newcommand{\FK}{{\mathfrak{K}}}
\newcommand{\mt}[1]{\textrm{\tiny #1}}
\newcommand{\CO}{{\cal O}}
\newcommand{\BR}{{\mathbb{R}}}

\begin{document}
\setlength{\unitlength}{1mm}

\title{Aspects of the bulk flat space limit in AdS/CFT}
\author{David Berenstein$^\dagger$ and Joan Sim\'on$^\ddagger$}
\affiliation {$^\dagger$ Department of Physics, University of California at Santa Barbara, CA 93106\\
$^\ddagger$ School of Mathematics and Maxwell Institute for Mathematical Sciences,\\
	University of Edinburgh, Edinburgh EH9 3FD, UK}

\begin{abstract} 
The flat space limit of scalar bulk fields in AdS is discussed within a Lorentzian canonical quantization setup tailored to describe AdS state preparation and to extract the flat S-matrix dynamics. We discuss how the algebraic \`{I}n\"on\"u-Wigner contraction captures the local physics of the equivalence principle in quantum field theory in a fixed background description. We develop the embedding formalism to describe the bulk AdS scalar primary wave functions as holomorphic functions. Flat space massive particle states are built out of the AdS primary together with AdS boosted wave functions. We compute their inner products and show that these become orthogonal in the flat limit,  resulting in the correct continuous spectrum for a standard unitary representation of the Lorentz group. In this same limit the original AdS descendants become null states. 
We also argue how the flat space S-matrix emerges from standard perturbation theory in the interaction picture.
To obtain flat space massless particles requires to consider a double scaled limit in which the boost rapidity is scaled to infinity keeping the average particle energy in the flat space limit fixed.  We comment on how this limit generates interesting massless state wave functions with non-trivial shape profiles  that remember the dimension of the AdS operator. We discuss some of the puzzles attached to these. 
\end{abstract}

\maketitle


\section{Introduction}

The AdS/CFT correspondence \cite{Maldacena:1997re} is a remarkable duality between quantum field theories and quantum gravity. Basically stated, the dual CFT can be used as a definition of quantum gravity, in asymptotically AdS spaces. It is incumbent on us to find how geometry and locality in higher dimensions arises from the CFT dynamics. This is where the conjectures on the gap of anomalous dimensions arise from \cite{Heemskerk:2009pn}. These can be used to prove from CFT assumptions that there is an effective  dual effective theory in the bulk \cite{Caron-Huot:2021enk}.

A natural notion of locality is that physics on scales much shorter than the AdS radius should be given by the physics of quantum gravity (or string theory) in flat space. It is understood that this description should be like the scattering S-matrix in flat space \cite{Polchinski:1999ry,Susskind:1998vk}. An S-matrix should be thought of as a real time process in a Lorentzian geometry. The procedure of getting to this natural statement that arises from the equivalence principle is called the flat space limit. 
Our goal in this paper is to explain some of the kinematics of this limit in the bulk in a simpler way than what has been done before.

The standard way to approach calculations in AdS space and their dual CFT correlators utilizes an Euclidean setup with the GKPW dictionary \cite{Gubser:1998bc,Witten:1998qj}. 
This obscures the Lorentzian physics required to understand small regions of spacetime, as well as the notion of in and out states. Many times these calculations are then analytically continued to try to get to Lorentzian physics. For a recent example see \cite{Marotta:2024sce},
where a 3 vector interaction vertex is treated in detail and it is argued how the flat space S-matrix arises from an analytic continuation of the Euclidean answer that they compute. 

A way to state the problem is as follows:  what is the correct preparation of states to get the locally flat physics? At least in some setups this is considered problematic \cite{Giddings:1999jq}. The other way to set up the problem is to begin this study by constructing a notion of a local bulk field. This is built by some integral representation on the boundary: the HKLL construction \cite{Hamilton:2006az} and their earlier avatars \cite{Banks:1998dd}. This procedure is problematic when interactions are included, but it is considered as a way to address the issue.

A third way to set up calculations is to start by analyzing Lorentzian wave functions in the bulk and then figuring out how they relate to the boundary at a later step. This is the route we follow in this paper. In that sense, we set up the problem of scattering as a question of determining the wave functions of incoming and outgoing states and then using time dependent perturbation theory to compute an amplitude. In presentation, this will be akin to a problem in canonical quantization rather than a problem in a path integral formulation of quantum field theory.

The main observation in our paper is that the flat space limit needs the following ingredients. A point $p$ in the bulk around which one will take a local flat space limit. The observation that the local Poincar\'e group at the point $p$ is a particular \`In\"on\"u-Wigner contraction \cite{Inonu:1953sp} of the AdS symmetry group.
Because of this property, we can take the limit at the level of the representation theory of the AdS symmetry group. In that sense, we construct global wave functions for incoming and outgoing particles for the limit, rather than strictly local physics.

There is a technicality about representation theory of the conformal group (the AdS symmetry group) and its relation to the flat space limit of massive particles that is essential for our work. For massive states in the flat space limit, the concept of primary states of AdS survives, becoming a massive particle at rest.  On the other hand, the descendants become null in the correct double scaling limit that leads to the Poincar\'e group. To recover the correct representation theory we have to work with primaries and their boosted states. These boosted states should be interpreted as a type of coherent state which becomes orthogonal to the primary in the flat space limit. The boosted states we construct end up providing exactly the particles at fixed momentum in the flat space limit and because of peculiarities of how they are constructed we can understand readily how they are prepared on the boundary\footnote{{There is a broad literature studying the representation theory of the conformal group, the Poincar\'e group and its carrollian avatars. We would like to point out \cite{Iacobacci:2024laa}, where the authors explain how to decompose massless and massive particles into unitary irreducible representations of the Lorentz group, for arbitrary spin, yielding the Mellin transform that appears in celestial holography. Also, \cite{Nguyen:2023vfz} builds conformal field representations of the Poincar\'e group living at null infinity (\`a la Mack-Salam) and shows how this relates to massless particles via Fourier-Mellin transforms.}}. 

The global wave functions that we produce for what become massive scalar particles in the flat space limit end up localized at $p$ on distances that are smaller than the AdS scale, which is also true of the interaction amplitudes we compute from these wave packets. Hence the flat space limit and the corresponding S-matrix is the appropriate description of the dynamics.

For massless particles, the flat limit is more subtle. A massless particle at rest makes no sense, so even the notion of primary states does not survive the limit. We can start instead with plane waves in the flat space limit and try to build global wave functions that produce such states, or we can just boost the primaries sufficiently in order to be able to survive the limit with finite energy.
In trying to imitate this boosted construction for massive particles for what become massless particles in the flat space limit we end up finding wave functions that are not fixed momentum states and the limit becomes very interesting.  For example, the wave functions remember the dimension of the operator in the shape of the wave packets in the flat space limit. This is in the kinematics of state preparation, not the dynamics.
The limiting wave functions end up being simple enough that we can exactly write their form as a superposition of plane wave states and they end up having some similarities with wave forms that appear in the celestial amplitude program.

The other observation we provide is that finding all these wave functions becomes extraordinarily simple in the embedding space formalism for Lorentzian AdS spacetimes. 
In a certain sense, this development alone is rather important, as we have not found a similar construction in the literature.

These states can then be easily assembled into integral representations of scattering amplitudes that are very suggestive of relations to other formulations of the flat space limit, not just in the strict AdS/CFT, but also for a (gapped) QFT in the flat space limit of curved AdS.  An example of such works can be found in \cite{Penedones:2010ue,Paulos:2016fap,vanRees:2022zmr} (see also the Snowmass white paper \cite{Kruczenski:2022lot}). 

A different formal issue about the limit is that the way we describe our approach to the problems above is by doing quantum field theory in a fixed curved spacetime. We just modify parameters in such a setup to describe what the limit must look like.
When using the AdS/CFT correspondence to infer  the dynamics of flat space physics through the so called \textit{flat limit}, one learns the latter takes place within a \textit{family of theories} in the boundary. The reason is as follows. Consider ${\cal N}=4$ SYM on $\mathbb{R}\times \text{S}^3$ to be concrete. One way to take the flat space limit keeps the string scale $\ell_s$ fixed and the string coupling constant fixed and small, while sending the ratio $\ell_s/R_{\text{AdS}}\to 0$ in order to make the AdS curvature vanish. Since the string coupling $g_s$ remains fixed, the rank $N$ of the gauge group $\SU(N)$ must satisfy $N\to \infty$. Since each value of N defines a different theory, we conclude that, technically, the flat limit is \textit{not} a limit in one conformal field theory, but rather a limit on a family of conformal field theories. We are not aware of any string theory setups where this conclusion is modified. What this means is that it is not clear that the limit respects the CFT axioms, and there might be some properties that we are implicitly assuming that might be violated. We do not currently have any deeper insight into this problem. The other point that is important to note is that in this paper we are describing scalar field scattering for simplicity. Our goal is to show that the flat limit is very natural.  We do not take into account that the AdS/CFT correspondence demands a theory of gravity in the bulk with gravitons, etc and the corresponding problem of building spinning wave functions as well. Such work is beyond the scope of the present paper and is being considered independently. 

Before finishing this introduction, we would like to explicitly mention the existence of a large literature discussing the flat limit of the AdS/CFT correspondence, both from the bulk and from the boundary perspectives. For a review on the carrollian structure emerging in the boundary theory, we refer the readers to \cite{Bagchi:2025vri}. For reviews on celestial holography, see \cite{Raclariu:2021zjz,Pasterski:2021rjz}. For a non-exhaustive list of references discussing bulk aspects of the limit, as in this manuscript, see \cite{Fitzpatrick:2010zm,Compere:2019bua,Compere:2020lrt,Campoleoni:2023fug,Bekaert:2024itn,Marotta:2024sce,Alday:2024yyj,Lipstein:2025jfj} and references therein.

This manuscript is organized as follows. In section \ref{sec:WI}, we motivate the standard \`In\"on\"u-Wigner algebraic contraction relating the conformal group to the Poincar\'e group from local bulk physics considerations. In section \ref{sec:ES}, we review the embedding formalism to describe AdS and explicitly apply it in section \ref{sec:swf} to describe scalar wave functions of AdS primary states as holomorphic functions. In section \ref{sec:Flat}, we study the flat space limit of AdS primary states giving rise to massive particles. We clarify the null state fate of the descendants, and show how AdS boosted wave functions become orthogonal, in the limit, filling the relevant massive particle multiplet.  In section \ref{sec:Amplitudes}, we describe the emergence of a flat S-matrix using canonical QFT methods in AdS. We also stress how our Lorentzian wave functions can be easily prepared by operator insertions in the AdS boundary. In section \ref{sec:massless}, we extend our analysis to massless particles, stressing some of the open questions that remain in this case. We finish with a summary of our results and a discussion on future research directions in section \ref{sec:future}.

\section{The \`In\"on\"u-Wigner contraction and the flat space limit}\label{sec:WI}

The Poincar\'e group is one of the possible \`In\"on\"u-Wigner contractions of the Conformal group of symmetries in AdS space, in the same sense than the group rigid translations of the plane is the \`In\"on\"u-Wigner contraction of the group of symmetries of a sphere \footnote{Similarly, the Galilei group and the Carrollian group are \`In\"on\"u-Wigner contractions of the Poincar\'e group \cite{levy1965nouvelle}.}. 
Our goal is to explain in what sense this limit can be taken at the level of the representation theory of the group. For that, we need two elements: the concept of the \`In\"on\"u-Wigner contraction as a limit and the proper representation theory of the group that we take a limit on. In this section we deal with the first : the group theory limit (the Lie theory of the appropriate vector fields in AdS). 

One of the fundamental aspects of General Relativity is the notion of the equivalence principle: the existence of local frames where the physics is essentially the physics of flat space. The existence of these frames can be stated as the existence of a coordinate system where at a spacetime point $p$ the Christoffel symbols vanish and the metric is locally constant. 

In these local frames, there is a notion of (local) momentum and  (local) Lorentz boost symmetry.
We will call the local coordinates at $p$, $x^\mu$. We also usually state that $g_p\equiv \eta_{\mu\nu}$ is the flat space metric, so that the $x^\mu$ behave like regular coordinates in Minkowski space. These local coordinates determine a frame at $p$.  
If we do a classical scattering experiment at the spacetime point $p$ (meaning the classical collision took place at the spacetime point $p$, at fixed position and time), the information of the incoming and outgoing particles which collided give rise to momentum vectors
$p^\mu$ in the tangent space at $p$ that are conserved. That is
\begin{equation}
\sum_{\text{in}} p^\mu= \sum _{\text{out}} p^\mu.
\end{equation}

We can also assume that after a choice of local time, all particles have positive energy and mass $p^0>0$.
The quantities $p^\mu \simeq dx^\mu \cdot(\partial_\tau)$ arise from evaluating the tangent vector to the curve of the incoming particle in the local coordinates.  Here $\tau$ is the proper time of the trajectory.

Similarly, there are local vector fields that generate Lorentz transformations at $p$ which act on the $x$ coordinates by a linear transformation and keep $p$ fixed. These preserve the local frame condition.
One can also include the vector fields $\partial_\mu \simeq \eta_{\mu\nu}(dx^\nu)$ as diffeomorphisms that move the frame to a nearby point. Since to first order $\partial_\mu g =0$, they are also local (flat space) translations. This usually fails at the next order: this is the order where curvature effects arise. All we need to ensure that the physics is that of flat space is that the scattering experiment takes place on regions that are much smaller than the curvature scale.
The important point is that from the point of view of classical physics, there is absolutely nothing mysterious here. 

Things change when we consider quantum fields. A lot of the structure of quantum fields in curved space is motivated by the idea that the physics should be approximately in a vacuum locally around each point $p$ that we choose. Unlike point classical particles, due to the uncertainty principle, quantum particles act as waves and understanding the physics near a point $p$ involves some global aspects, not just purely local information. The problem goes away when we consider wave lengths that are much shorter than the curvature scale. Such wave packets can be localized to very small regions and curvature can be ignored. Such physics is associated to the UV degrees of freedom of the field theory.
The idea of a flat space limit is to take a double scaling limit
where the wavelengths go to zero and the units are rescaled so that the local Lorentz physics is the correct answer to scattering problems in wave mechanics. This is schematically written as 
\begin{eqnarray}
  \frac{\lambda}{R} &\to& 0, \nonumber \\
   \braket {\text{out}}{\text{in}}&\longrightarrow\atop{\lambda\to 0} &   \  S\ \delta(\sum p_{\text{in}}-\sum p_{\text{out}}).
\label{eq:bulk-flat}
\end{eqnarray}
where $S$ is a rescaled version of the $S$-matrix appropriate to the flat space limit.

Let us consider the discussion above when the spacetime geometry is AdS. Since AdS is a maximally symmetric spacetime, it has a group of symmetries with the same number of generators as the local Poincar\'e group. Given an AdS point $p$, we can identify the Local Lorentz group generators with the subset of AdS Killing vectors vanishing at $p$, i.e. leaving the point $p$ fixed. The rest of the generators must be mapped into the local translations.

The above argument identifies two types of generators. The Lorentz group generators $L_{\mu\nu}$ and the translations $P_\mu$ of the local Poincar\'e algebra. These satisfy the schematic commutation relations
\begin{equation}
    [L,L] \propto L\,, \qquad [L,P] \propto P\,, \qquad  [P,P] =0
\label{eq:MinIso}
\end{equation}
The same decomposition can be done with the AdS vector fields. They satisfy almost the same algebra
\begin{equation}
    [L,L] \propto L\,, \qquad  [L,\tilde P] \propto \tilde P\,, \qquad [\tilde P,\tilde P] \propto L 
\label{eq:AdSIso}
\end{equation}
except that now the \textit{translations} $\tilde P$ do not commute.

The two notions of $L$'s are the same, but the corresponding $P$'s and $\tilde P$'s are different because they have different commutators. The idea of the \"In\"on\"u-Wigner contraction is that in the same way that we take the wavelength to zero near $p$, we will need to rescale the $\tilde P$ generators of AdS to land on a notion where $P$ has a finite value: the flat space limit notion of momentum.

Essentially, we introduce a new parameter $\Lambda$ associated to a scale\footnote{One can think of $\Lambda=R_\text{AdS}$ as the AdS radius.}, and we set 
\begin{equation}
    P = \frac{\tilde P}{\Lambda}
\label{eq:P-scaling}    
\end{equation}
The commutators of the $P$ are then given by
\begin{equation}
    [P,P]= \frac 1{\Lambda^2} [\tilde P, \tilde P]
\end{equation}
At this stage, we have done nothing more than to rescale $\tilde P$. When we take $\Lambda\to \infty$ with fixed $P$, we implement the precise double scaling limit that we alluded to earlier in \eqref{eq:bulk-flat}.
Then the limit of the last commutator in \eqref{eq:AdSIso} becomes the last commutator in \eqref{eq:MinIso}. This limit procedure is called the \`In\"on\"u-Wigner contraction. 

The double scaling of \eqref{eq:bulk-flat} that we want is therefore thought of as figuring out the correct in and out states that permit one to fix the kinematics of the initial and final states of the S-matrix as follows
\begin{eqnarray}
 \Lambda \simeq 1/\lambda &\to& \infty, \nonumber \\
\ket{\text{in}}&\longrightarrow\atop{\Lambda\to \infty} &  \ket{p_{\text{in}}}\\
\ket{\text{out}}  &\longrightarrow\atop{\Lambda\to \infty} & \ket{p_{\text{out}}}
\label{eq:prep_states}
\end{eqnarray}
where the states on the right hand side are built in flat space QFT with very small uncertainty on the momentum, while the states on the left are global wave functions and the $\braket{\text{out}}{\text{in}}$ amplitudes are amplitudes for global AdS states that coincide with what the right hand side would demand.

Operationally, we think of $\tilde P$ in a semi-classical limit. Given some semi-classical wave packet with an expectation value of the generators $\tilde P$ given by a number with the same name $\tilde P$ and a small dispersion, the double scaling limit involves taking a family of wave packets and scaling their semi-classical quantum numbers while keeping the dispersion small. In the limit, if the ratio of the dispersion to $\tilde P$ can be taken to be vanishingly small, we end up landing in states with fixed $P$ quantum numbers: the notion of a momentum eigenstate.  

Once the limit is phrased in terms of wave packets rather than an abstract Lie algebra, the limit is also implicitly taken as a limit in the representation theory of the AdS isometries: representations of the conformal group (since $\text{AdS}_{d+1} = \text{Conf}_d)$. Part of our goal is to understand how this process works at the level of representation theory.

To that effect, consider the usual algebra of the conformal group. This is depicted in table \ref{tab:Conf}. The table is organized by the conformal weight of the generators. 
\begin{figure}[ht]
\begin{tabular}{|c|c|c|}
\hline
 Conformal Symmetry generator    &  Notation & Conformal weight \\
 \hline
Dilatation (energy)     & $\Delta$ & 0\\
Rotations & $R^i_j$& 0\\
Conformal translations & $\vec {\mathfrak P}$& 1\\
Special conformal generator & $\vec {\mathfrak K}$& -1\\
\hline
\end{tabular}
    \caption{Conformal group generators} \label{tab:Conf}
\end{figure}

The generators are classified as follows. We have the dilatation operator $\Delta$, the rotation generators $R^i_j$, the conformal translations  $\FP$ and the special conformal generators $\FK$. These translations $\FP$ are different than $\tilde P$ above, so we need to use a different notation for them. Usually these act on a Euclidean plane. There is a conformal rescaling that takes us to the cylinder and radial time becomes the time on the cylinder. This is the time that we identify with global time in the CFT.  We build a Lorentzian model by analytically continuing this radial time.
The translation generator acts as a raising operator and increases the conformal weight by one.
Similarly, the adjoint ${\mathfrak K}= {\mathfrak P}^\dagger $ that corresponds to the special conformal generator is a lowering operator and has weight $-1$. 

The global time of the CFT can be extended to a full time coordinate in $AdS$ such that $\partial_t$ is an isometry and the time slices are orthogonal to $\partial_t$. Such a vector field $\partial_t$ is nowhere vanishing. 
There is a special point in $AdS$ at fixed time which is invariant under the rotations $R$. This is called the origin of $AdS$. We will choose our flat space limit so that it coincides with double scaling around the chosen point $p$ at time zero, which is at the origin.

\subsection{Taking the limit}

We have fixed a point $p$. Now we can proceed with classifying the different vector fields according to how they act in the vicinity of $p$ to distinguish local Lorentz rotations versus translations. 
Since the rotations leave the point $p$ invariant, they become part of the local Lorentz group. 
By contrast, since $\partial_t$ is nowhere vanishing, it must become one of the $\tilde P$ generators. 
In particular, we distinguish it as the energy as it is a timelike Killing vector field.
The generator $\Delta$ becomes $\tilde E$. That leaves us with mapping $\FK$ and $\FP$ to the rest of the generators of the flat space isometry group. Since both of these transform as vectors under rotations,  one linear combination becomes $\vec {\tilde P}$ and the other linear combination becomes the boosts $\vec B$ which are part of the local Lorentz group. Hermiticity of $\vec P, \vec B$ requires that the linear combinations are Hermitian. If $\vec {\tilde P} = \alpha \vec\FP+\gamma \vec \FK$, then $\gamma=\alpha^*$ are complex conjugates of each other. Similarly,  we have  $\vec{ B} =  \beta \vec\FP+\beta^* \vec \FK$

Eventually we need to determine these linear combinations precisely.

What is important is that $\tilde P$ is rescaled, but $\vec B$ is not. Hence, the limit does not work nicely directly in terms of the generators $\FK,\FP$.
Since the representation theory of the conformal group is built precisely by actions of $\FP$ on primary states, that are determined by how $\FK$ acts on states, this means that the limit has to be taken in a non-trivial way to get something meaningful. The details of what we have written above can be found in the table \ref{tab:map}.

\begin{figure}[ht]
\begin{tabular}{|c|c|c|}
\hline Symmetry generator & AdS Symmetry group & Flat space limit \\
\hline Energy & $\tilde E= \Delta$ & $E= \Delta/\Lambda$ \\
Rotations
     &  $R^i_j$& $R^i_j$\\
Momentum & $\vec{\tilde P}= \alpha \vec {\mathfrak P}+ \alpha^*\vec {\mathfrak K}$ & $\vec P= \vec {\tilde P}/\Lambda$    \\
Boosts & $\vec B = \beta \vec {\mathfrak P}+  \beta^*  \vec{\mathfrak K}$ & $\vec B$\\
\hline
\end{tabular}
\caption{How the CFT generators are mapped to flat space }\label{tab:map}
\end{figure}

What we can say with confidence is the following. The representations of the conformal group are built by diagonalizing $\Delta$. For each irreducible representation, there is a lowest energy state called the primary, which is annihilated by the $\FK$
\begin{equation}
    \hat \Delta | \psi_{\Delta} \rangle=  \Delta\,|\psi_\Delta\rangle\,, \quad \FK^i |\psi_\Delta\rangle = 0\,.
\label{eq:primary}
\end{equation}
Due to the algebra \eqref{eq:AdSIso}, the energy splitting between members of the same multiplet is quantized. The usual convention is for this splitting to be in multiples of one, which is the dimension of $\FP$.
The other states of the multiplet, called descendants, are produced by acting with polynomials of $\FP$ on the primary state. Since the primary state $|\psi_\Delta\rangle$ is the minimum energy state over all states of the representation, it can also be assumed that it has the minimal energy of all the states in the \`In\"on\"u-Wigner limit.

Since the energy $\Delta$ is one of the AdS translations $\tilde{\text{P}}$, it must be rescaled to identify the flat space energy $E=\frac{\Delta}{\Lambda}$. We learn this way that finite mass particles in the flat space limit correspond to operators of parametrically large dimension $\Delta$. The spectrum will become continuous as the splittings in dimension by units of $1$ from the descendants rescales to zero. In particular, within members of the conformal group multiplet, the splitting in $\Delta$ between states is of order one. Hence, descendants will have the same energy as the primaries in the flat limit. However, since they must also be orthogonal to the primary state, this suggests the notion of descendant breaks down in the \`In\"on\"u-Wigner limit. We will explain this point in detail below.

This conclusion can be traced back to the fact that $\FK, \FP$ must be a linear combination of $\tilde P, \vec B$.
Since $\tilde P$ must be scaled, it is natural to rescale $\FP$ so that its action becomes degenerate in the limit. Essentially, $\FP/\Lambda \propto \tilde P/\Lambda+O(1/\Lambda) \simeq \FK/\Lambda +O(1/\Lambda)$ so when we take the limit of the generators $\FK,\FP$, they become indistinguishable. Since $\FK$ acts by zero, descendants become null states. We will make this very explicit  when we consider the double scaling limit in the embedding space formalism and describe it correctly using the local coordiates that arise from that construction.

The upshot of this discussion is that we can not follow the path of the primary/descendant representations of the conformal group when attempting to understand the \`In\"on\"u-Wigner contraction at the level of the representation theory. One of the goals in this paper is to explain the correct way to take the limit of representations to recover the correct flat space limit of relativistic QFT.

\section{The embedding space formalism}\label{sec:ES}

We will consider the flat space limit in the embedding space formalism.
We are interested in the Lorentzian version of this problem, rather than the Euclidean version which has been developed in \cite{Costa:2011mg,Costa:2014kfa}.

$\text{AdS}_{d+1}$ is the (universal cover of the) hyperboloid\footnote{Notice we set $R_\text{AdS} =1$.}
\begin{equation}
  \eta_{\text{AB}} Y^\text{A} Y^\text{B} = -1 \quad \Leftrightarrow \quad 1= (y^{(-1)})^2+(y^0)^2-\vec y^2
\label{eq:AdS-hyp}
\end{equation}
in the embedding space $\BR^{2,d}$ with coordinates $Y^A = \left(y^{-1}, y^0,\vec y\right)$, where $\vec y = (y^1,\dots , y^d)$. For later reference, we shall refer to the point ${\bf 0}$
with coordinates $y^{(-1)}=1,y^0=0, \vec y =0$ as the origin in $\text{AdS}_{d+1}$.The $\text{AdS}_{d+1}$ metric is induced from the flat space metric in the embedding space $\BR^{2,d}$ \begin{equation}
  ds^2= \eta_{\text{A}\text{B}} dY^\text{A} dY^\text{B} =-(dy^{(-1)})^2-(dy^0)^2+ (d\vec y)^2\,.
\label{eq:emb-metric}
\end{equation}

The advantage of the coordinates $Y^A$ is that they transform linearly under $\SO(2,d)$, the $\text{AdS}_{d+1}$ isometry group. It is convenient to introduce the complex coordinates
\begin{equation}
z= y^{(-1)}+ i y^0
\end{equation}
so that the embedding equation \eqref{eq:AdS-hyp} becomes
\begin{equation}
1= z \bar z- \vec y^2 
\end{equation}
It follows that any point in $\text{AdS}_{p+1}$ requires $|z|\geq 1$. Thus, the holomorphic coordinate $z$ is defined on the complex plane with a disk of size one removed. Topologically, this describes a non-simply connected space.

Introducing polar coordinates and an angular decomposition 
\begin{equation}
  y^{(-1)}= \cosh\rho\,\cos(t)\,, \qquad y^0= \cosh \rho\, \sin(t)\,, \qquad  \vec y = \sinh \rho\, \vec n\,,
\label{eq:global-coord}
\end{equation}
in terms of the unit vector $\vec n$, the $\text{AdS}_{d+1}$ metric \eqref{eq:emb-metric} becomes
\begin{equation}
ds^2= - \cosh(\rho)^2 dt^2 + d\rho^2 +\sinh(\rho)^2 d \Omega_{d-1}^2
\label{eq:global-AdS}
\end{equation}
where $d\Omega_{d-1}^2$ is the metric of a round sphere, $\rho$ is the radial AdS coordinate and $t$ is the global AdS time. The AdS boundary is reached by $\rho\to \infty$. It is a cylinder parameterized by time $t$ and the unit sphere $\vec n$, with metric
\begin{equation}
    ds^2_{\partial \text{AdS}}= \lim_{\rho \to\infty} \exp(-2 \rho) ds^2|_\rho= -dt^2+d\Omega_{d-1}^2
\label{eq:AdS-boundary}
\end{equation}

By definition, the bulk metric \eqref{eq:global-AdS} is $\SO(2,d)$ invariant. Since $\partial_t$ is an everywhere timelike Killing vector field of this metric, the energy operator $\Delta$ must be identified with $i \partial_t$. The set of AdS isometry generators is given by
\begin{equation}
\begin{aligned}
  R^i_j &= y^i\partial_{y^j} -y^j \partial_{y^i} \\
  \Delta &= i\partial_t = \bar z \partial_{\bar z}-z \partial_z. \\
  \FK_i &= z\partial_{y^i} + 2 y^i \partial_{\bar z}\\
  \FP_i &=  \bar z\partial_{y^i} + 2 y^i \partial_{ z}
\end{aligned}
\end{equation}
where $R$ are rotations of the vector $\vec n$. The action of $\Delta$ and $R$ as time translations and rotations, respectively, is manifest in the AdS boundary metric \eqref{eq:AdS-boundary}.
Notice that $[\vec \FK,\vec \FK]=0=[\vec \FP,\vec \FP]$. Furthermore,  since $\Delta$ acts like the dilatation operator, the generators $\FK$ and $\FP$ have fixed weights with respect to $\Delta$ 
\begin{equation}
  \left[\Delta,\,\FP_i\right] = \FP_i\,, \qquad \qquad  \left[\Delta,\,\FK_i\right] = -\FK_i\,.
\end{equation}
It follows $\FK$ lowers the energy while $\FP$ increases it. Notice these statements follow from the nice action of the operator $\Delta$ on $z$ and $\bar z$, since $z= \cosh(\rho) \exp(i t)$. 

The vector fields $R,\Delta, \FK, \FP$ are defined on the embedding space. They are equivalent to $\eta_{\text{A}\text{C}} Y^\text{C}\partial_\text{B} - \eta_{\text{B}\text{C}} Y^\text{C}\partial_\text{A}$, making their linear action manifest. We view them as extensions of vector fields in AdS space with the same names. Since they leave the hyperboloid \eqref{eq:AdS-hyp} invariant, when restricted to AdS, they produce flows that remain in the AdS hyperboloid.

\subsection{Scalar wave functions}\label{sec:swf}

AdS wave functions can be constructed using the representation theory of the conformal group in terms of a primary field and its descendants. As indicated in \eqref{eq:primary}, the primary field has positive energy $E= \Delta$ and is annihilated by $\FK^i$. Hence, its wave function $\psi_\Delta$ satisfies 
\begin{equation}
    \hat H \psi_{\Delta}= i \partial_t \psi_{\Delta} = (\bar z \partial_{\bar z}-z\partial_z) \psi_{\Delta} = \Delta\,\psi_\Delta\,, \quad \FK^i\psi_\Delta = \left(z \partial_{y^i} + 2y^i\partial_{\bar z}\right)\psi_\Delta= 0\,.
\end{equation}
For scalar fields, the primary field is rotationally invariant. Thus, its wave functions can only depend on $t$ and $\rho$. Within the embedding space, we can reach an equivalent description. Since $\psi_\Delta$ is independent of the $y^i$, it follows $\psi_\Delta(z,\bar z)$. The two remaining and defining equations reduce to
\begin{equation}
  \hat\Delta \psi_\Delta =  (\bar z\partial_{\bar z}- z \partial_z) \psi_\Delta = \Delta\,\psi_\Delta\,, \qquad \vec \FK^i \psi_\Delta \propto \partial_{\bar z} \psi_\Delta = 0
\end{equation}
The general wave function for scalar primary fields is the holomorphic function
\begin{equation}
    \psi_{\Delta}= \frac 1{z^\Delta} = \frac{1}{[\cosh(\rho) \exp(i t)]^\Delta}\,.
 \label{eq:AdS-primary}  
\end{equation}
The wave function on the left is implicitly living in ${\BR}^{2,d}$, whereas the one on the right is strictly in AdS. They coincide on the hyperboloid. 
As a check, one can easily verify that the wave function \eqref{eq:AdS-primary} solves the radial bulk wave equation
\begin{equation}
- G^{-1}(\rho) \partial_\rho \left(G(\rho) \partial_\rho \psi_{\Delta}\right) -\frac{\Delta^2}{\cosh(\rho)^2} \psi_{\Delta} -\Delta(\Delta-d) \psi_\Delta
=0
\end{equation}
arising from an AdS scalar field of mass squared $m^2= \Delta(\Delta-d)$ and frequency $\Delta$. Notice $G(\rho)= \cosh\rho\,(\sinh \rho)^{d-1}$ is the AdS volume form $\sqrt{-\det g}$. Mathematically, the wave function $z^{-\Delta}$ is holomorphic (harmonic) outside the disk $|z|\geq 1$. It actually lives in the universal cover of the AdS hyperboloid as $\Delta$ is a real number. It would have been multi-valued otherwise. Notice how this construction simplifies the analysis relative to other works (for example \cite{Fitzpatrick:2010zm}, which uses a different coordinate system).

Two remarks will help us to develop some semi-classical interpretation for the wave functions \eqref{eq:AdS-primary}. First, notice that the radial profile of $\psi_\Delta$ is approximately Gaussian in $\rho$ around zero for large $\Delta$ since
\begin{equation}
    \left | \psi_{\Delta} \right| = \frac{1}{(\cosh \rho)^\Delta}\simeq \frac 1{(1+\rho^2/2)^\Delta}\,.
    \simeq \exp(-\Delta \rho^2/2)
\label{eq:width}
\end{equation}
Thus, the wave function becomes negligible when $\rho \gg 1/\sqrt{\Delta}$. In our units, a distance of order one in $\rho$ is order one in the radius of AdS. Hence for large $\Delta$, the wave function is localized on distances much shorter than the AdS scale. Indeed, if $\Delta$ is large, one can think of a particle at the bottom of AdS as a particle in a harmonic oscillator trap \cite{Berenstein:2002jq}, whose ground state wave function is Gaussian.

Second, notice that the wave function is peaked at $\rho\simeq 0$ for all time. That locus is the timelike geodesic sitting at $\rho=0$ and parametrised by $t$. We can think of it as a point particle at rest, sitting at the center of AdS, namely $\rho=0$. Because of the AdS curvature, it does not get dispersed in time.

Primary wave functions of large $\Delta$ can be thought of as (semi) classical particles of mass $m_{AdS}$ in the minimal energy allowed: a particle at rest at $\rho=0$. In such a limit, the energy of the particle $\Delta$ is roughly $m_{AdS}+d/2$. The factor of $d/2$ can be interpreted as a zero point energy from thinking of AdS as a harmonic trap (see for example \cite{Berenstein:2002ke}).

On the other hand, if $\Delta \simeq O(1)$, the wave function behaves asymptotically as $\exp(-\Delta \rho)$ and is supported on a radius of order one, i.e. the radius of the wave packet is the size of the AdS radius itself and the interpretation of the system as a classical point particle doesn't make much sense.

Given a primary state $|\psi_\Delta\rangle$, descendants are generated by acting multiple times with $\FP$. If $\FP$ acts $s$ times, the resulting wave functions look as follows
\begin{equation}
\psi_{\text{desc}} \equiv z^{-(\Delta+s)}\,f_s( \bar z z, \vec y)\,,
\end{equation}
where $f_s$ is a polynomial of degree $s$ in the combined variables.  The $\psi_{\text{desc}}$ are (generalized) harmonic functions in the embedding space ${\mathbb R}^{2,d}$ and carry energy $E= \Delta +s$. This is a generalization to AdS of the construction of spherical harmonics as harmonic polynomials in an embedding space (see the discussion in \cite{vanNieuwenhuizen:2012zk} and references therein, which we found useful). We can then restrict to the AdS hyperboloid and rewrite these in terms of 
the $\rho, t, \Omega$ variables if we want to.

Given a primary and the descendants, we have a representation of the conformal group on scalar waves of AdS. We will think of these as generating single particle states in the bulk. We can do quantum field theory in curved space with these wave functions.  

For each descendant, one associates a lowering operator for the quantum field on AdS. Consider a bulk time slice at $t=0$. Incoming wave functions are taken to have positive energy, so they are being measured by lowering operators. For outgoing wave functions, one needs to complex conjugate. This produces negative energy states that are annihilated by  $\FP=\FK^\dagger$ rather than by $\FK$. These are assigned to raising operators rather than lowering operators and their wave form results in outgoing wave functions (wave functions that come from the future to the interaction point instead).

Once the wave functions of the entire bulk conformal multiplet have been constructed, let us describe their normalizability. Consider the primary wave functions \eqref{eq:AdS-primary}. Using the Klein-Gordon norm evaluated on the global AdS time slice $t=0$
\begin{eqnarray}
\langle \Delta | \Delta\rangle & \equiv& -\int_{t=0} \sqrt{-g} g^{tt} \psi_\Delta^* i{  \overset{\leftrightarrow} {\partial_t}} \psi_\Delta\\
&=& 2\Delta\,\text{vol}(\text{S}^{d-1}) \int d\rho \cosh(\rho)^{-1}\sinh(\rho)^{d-1} |z|^{-2\Delta} = 2 \Delta\,\text{vol}(\text{S}^{d-1})\, \frac{\Gamma[d/2]\Gamma[\Delta+1-d/2]}{\Gamma[1+\Delta]}>0
\label{eq:norm}
\end{eqnarray}
This converges if $\Delta>d/2-1$. The pole in the Gamma function takes care of the domain of convergence by requiring that the norm be positive. The constraint is that for a propagating field in AdS satisfies $\Delta> d/2-1$. The special case $\Delta=d/2-1$ gives rise to the singleton representation that is not supported in the bulk of AdS, but only on the boundary. Basically, this says that normalizability of the primary wave function in AdS is equivalent to the unitarity bound in conformal field theory.

\section{The flat space limit in the embedding space formalism and representation theory}\label{sec:Flat}

Having explained the embedding space formalism and identified the point $p= (1, 0, \vec 0)$ as the AdS origin, let us identify the local Lorentz group with the subset of AdS isometries leaving $p$ fixed. Since the Lorentz group acts linearly on the embedding coordinates $y^{-1}, y^0, \vec y$, this condition requires no action on $y^{-1}$. That is, the \textit{local} Lorentz generators are given by rotations and boosts
\begin{equation}
R^i_j = y^i \partial_{y^j}-y^j\partial_{y^i},\, \qquad B^i= y^i \partial_{y^0} +y^0 \partial_{y^i} \,.
\end{equation}
Since $y^0 = z - \bar z$, it follows
\begin{equation}
  B^i \propto \frac 1{2i} (\FK^i-\FP^i)\,.
\end{equation}
This fixes the specific linear combination of $\FK$ and $\FP$ giving rise to a Lorentz boost. The other linearly independent combination gives rise to the space translations
\begin{equation}
\tilde P^i \propto \frac 1{2} (\FK^i+\FP^i) \simeq y^{-1} \partial_{y^i}+y^i \partial_{y^{-1}}\,.
\end{equation}
These are indeed acting non-trivially on $y^{-1}$. Finally, and as discussed earlier, the energy $\tilde E = \tilde P^0 = y^{-1}\partial_{y^0}- y^0 \partial_{y^{-1}}$ is the one that mixes $y^{-1}$ and $y^0$.

In the global coordinates \eqref{eq:global-coord}, the point $p$ corresponds to $\rho=0$ and the choice of Cauchy time slice $t=0$. Let us now describe the physics near $p$ to implement the flat limit of equation \eqref{eq:bulk-flat}. First, since $y^{-1}\simeq 1$ and varies quadratically with the $y^i \simeq \rho\,n^i$, we can ignore derivatives with respect to $y^{-1}$ (since they are subleading near $p$). Thus near $p$, we have that $\tilde P_i \simeq \partial_{y^i}$. Similarly, $y^0\simeq \cosh(\rho) \sin(t) \simeq t$. That is, $y^0$ can also be interpreted as the global time coordinate in the flat space limit. Consequently, $\tilde E= \tilde P^0 = \partial_{y^0}$.

The rescaling of the generators \eqref{eq:P-scaling} $P^0=\tilde P^0/\Lambda$ and $P^i= \tilde P^i/\Lambda$ is equivalent to going to a new coordinate system given by rescaled variables
\begin{equation}
    y^{0,i} = \Lambda^{-1}\, x^{0,i}
\end{equation}
where $x$ is kept of order one. In that case $\partial_{y^{0,i}}=\Lambda \partial_{x^{0,i}} $ and dividing by $\Lambda$ gives us \textit{finite} momenta $P_\mu=\partial_{x^\mu}$ in flat space,
as we would expect in a flat coordinate system. Notice that in the limit, $y^{-1}\equiv 1+O(1/\Lambda^2)$.

The advantage of the embedding space formalism now becomes apparent. The flat space limit is relatively trivial in the embedding space coordinates. 
The scalar primary wave function \eqref{eq:AdS-primary} becomes
\begin{equation}
    \psi_{\Delta}=\frac 1{(1+i x^0/\Lambda)^\Delta}\simeq \exp(-i x^0 \Delta/\Lambda)= \exp(-i x^0 m)\,,
\end{equation}
whereas the primary condition $\FK_i \psi_\Delta=0$ reduces to $\partial_{x^i}  \psi_\Delta= 0$. Hence, the wave function must be translation invariant and at constant frequency. This is exactly the wave function of a massive particle at rest in the frame determined by a constant time slice.
The flat space mass $m$ is defined by $m= \lim_{\Lambda\to \infty} \Delta/\Lambda $, i.e. the rescaled energy of the particle in the $x$ coordinates. 

Conformal multiplets are generated by acting with $\FP^i$ on the primary state. However, in the flat limit $\FP^i \simeq \Lambda \partial_{x^i} $, so we need to act with $\FP^i/\Lambda$ rather than the $\FP^i$. These do \textit{not} produce new states as they become degenerate with the double scaled $\FK^i/\Lambda$.
So, how do we build the correct representations of quantum fields for the flat space limit? We need to think \`a la Wigner. 

Once the wave functions for a particle at rest are known, the wave functions for a moving particle are generated by boosting them (this is part of the standard foundations in QFT and it is especially well presented in \cite{Weinberg:1995mt}). This procedure amounts to a simple change of coordinates by a Lorentz transformation $(x^\mu)'= \Lambda^\mu_\nu x^\nu$. Once more, the embedding space formalism described earlier allows us to perform this boost globally in AdS, and not just locally at $p$. Also, since the action involves an AdS isometry, it preserves norms, i.e. it is a unitary transformation in the full Hilbert space of single particle states in AdS. This last point is crucial, as once the norm of the primary state is fixed, all other states obtained by rotation this way have a known norm. 
Via the AdS/CFT correspondence, this must be the same as the norm of the state in the dual CFT.

Following Wigner, we perform a boost of rapidity $\eta$ along the direction $\hat n$ so that our original embedding coordinate $z$ gets rotated to $\xi$
\begin{equation}
z\to \xi = y^{-1} + i( \cosh(\eta)y^0+\sinh(\eta) \hat n \cdot \vec y )\,.
\label{eq:xi}
\end{equation}
Introducing a unit (future pointing) time like vector $k^\mu=(\cosh\eta,\,-\sinh \eta\,\hat n^i)$ allows us to write a more covariant expression
\begin{equation}
 \xi = y^{-1} - i k_\mu y^\mu
\end{equation}
with $y^\mu=(y^0, \vec y)$. In fact, defining a (complex) null vector in $\BR^{2,d}$ with components $\kappa^\text{A}=(-1, -i k^\mu)$ extends the covariant form to the embedding space since $\xi= \kappa \cdot Y = \eta_{\text{A}\text{B}}\kappa^\text{A} Y^\text{B}$.

The boosted AdS wave function is given by 
\begin{equation}
    \psi_{\Delta,\kappa}= \frac{1}{ \xi^\Delta}\equiv \frac{1}{(\kappa \cdot Y)^\Delta}\,.
\label{eq:bwf}
\end{equation}
Some readers may recognise these wave forms as an analytic continuation of wave functions in Euclidean AdS \cite{Costa:2011mg}, where the notion of $\kappa$ has also been analytically continued.

To discuss scattering processes, we will also need the \textit{out} states, carrying negative energy. At rest, these are given by the complex conjugate wave functions
\begin{equation}
    \psi^*_{\Delta}= \frac 1{\bar z^\Delta}.
\end{equation}
In that case, to boost them we need a similar $\kappa$, but now $\Im m (\kappa) \simeq (0, k_\mu)$ is determined by a past pointing timeline unit vector instead. In this way, both positive energy and negative energy wave functions lead to very similar descriptions and the difference between $\bra{\text{out}}$ and $\ket{\text{in}}$ states can be blurred: this allows us to generalize $k$ to be a complex timelike unit vector $ k\cdot k=-1$ in ${\mathbb C}^{1,d}$ if necessary. This is what is required to analytically continue to Euclidean AdS space and to be able to transform in states into out states in Euclidean correlation functions (the idea of crossing symmetry). Basically, we can now talk about analyticity in variables that are analogous to momentum: analyticity in the variables $k$. 

Contrary to flat space physics, AdS boosted wave functions are \textit{not} orthogonal to primary states (this is why we do not use them to build conformal multiplets). However, we shall show how they become orthogonal in the flat limit. To compute their AdS inner products, notice the boosted embedding coordinate can be written as
\begin{equation}
 \xi = \frac{1}{2}(z +\bar z)+ i \cosh(\eta) \frac{1}{2i}(z-\bar z) + \dots = \frac{1}{2}(1+ \cosh \eta)z +\dots 
\end{equation}
where we used $y^{-1}= \frac{1}{2}(z +\bar z)$ and $y^0=\frac{1}{2i}(z-\bar z)$. Since $|z|>1$, the boosted AdS wave function can be Taylor expanded in powers of $1/z$ as follows
\begin{equation}
\frac{1}{ \xi^\Delta}= \frac {2^\Delta}{(1+\cosh(\eta))^\Delta}  \frac 1 {z^\Delta} +\dots = \left(\frac{1}{\cosh \frac{\eta}{2}}\right)^{2\Delta}  \frac 1 {z^\Delta}  + \dots
\end{equation}
due to the identity $\cosh^2 \frac{\eta}{2} = \frac{1}{2}(1+\cosh \eta) $ and where the additional terms come with powers of $1/z^{\Delta+1}$ or higher. Since these extra terms carry 
different energy to the original primary state, they are orthogonal to it. Finally, if the states with AdS wave function $\psi_\Delta$ are properly normalised (dividing by \eqref{eq:norm}), one finds
\begin{equation}
    \braket {\Delta, \kappa} {\Delta}=  \left(\frac{1}{\cosh \frac{\eta}{2}}\right)^{2\Delta} \,.
\label{eq:inner-1}
\end{equation}

This result confirms that AdS boosted wave functions are not orthogonal to primary states. Once more, one can write the inner product in covariant form in terms of $\kappa$. Indeed, noticing that $1+\cosh(\eta) = - \kappa^*\cdot  \kappa_0 $, where $\kappa^*$ is the complex conjugate vector to $\kappa$ and $\kappa_0$ is the vector associated to the original primary wave function $\psi_{\Delta}$, namely $\kappa_0= (-1, i, \vec 0)$,  the inner product equals
\begin{equation}
    \braket {\Delta, \kappa} {\Delta}= \frac {2^\Delta}{ (- \kappa^*\cdot  \kappa_0)^\Delta}\,.
\label{eq:inner-2}
\end{equation}
This allows to extend the above calculation to any pair of such states
\begin{equation}
    \braket {\Delta, \kappa} {\Delta,\kappa'}= \frac {2^\Delta}{ (- \kappa^*\cdot  \kappa')^\Delta}\,.
\label{eq:inner-3}
\end{equation}
It is reassuring to double check the Lorentz invariance of these inner products, since $\kappa^*\cdot  \kappa'$ is ${\mathbb R}^{2,d}$ invariant, which includes the invariance along the directions spanned in ${\mathbb R}^{1,d}$. The non-vanishing of the inner product \eqref{eq:inner-3} allows us to interpret AdS boosted wave functions as some kind of coherent state giving rise to an overcomplete basis.

Now, consider the flat space limit of the inner product \eqref{eq:inner-1} for any non-vanishing rapidity $\eta \neq 0$. Since $\cosh\frac{\eta}{2}>1$, it follows
\begin{equation}
    \braket {\Delta, \kappa} {\Delta}_{\text{flat}}= \lim_{\Delta=\Lambda m\to \infty} \left(\frac{1}{\cosh \frac{\eta}{2}}\right)^{2\Delta} =0.
\end{equation}
The same conclusion holds between the inner products of two AdS boosted wave functions with different rapidity using \eqref{eq:inner-3}. Thus, AdS boosted wave functions with a finite mass in the flat space limit become orthogonal to each other.

Let us close this discussion with a comment regarding the normalization of the resulting flat space wave functions. Since the spectrum of $E$ is continuous, this normalization involves
a $\delta$ function. One can understand the AdS origin for this normalization as follows. The flat space coordinates $x$ have an infinite extent. However, once $x$ reaches a scale of order $\Lambda$, the AdS coordinates $y$ become of order one. Since curvature effects become important, one must cut off the integral. In fact, since the width of the semi-classical wave packets scales like $1/\sqrt{\Delta}$ in \eqref{eq:width}, these curvature corrections arise sooner at a scale $\sqrt{\Lambda}$. We conclude the states have norm of size $\Lambda^{d/2}$, which becomes infinite in a double scaled version where we forget $\Lambda$.

\section{Amplitudes and state preparation}\label{sec:Amplitudes}

We are now ready to define amplitudes in AdS from the wave functions we have constructed so that the amplitude is dominated by the physics of a collision at $y^0=0, \vec y =0$, i.e. in flat space. We aim to stress the relevance of state preparation in AdS, its consistency with the insertion of boundary operators and their normalization, and how a notion of an S-matrix emerges in the flat limit.

The idea of an amplitude is to set it up in time dependent perturbation theory for quantum field in curved space assuming that it is like a finite box. Since AdS has a global timelike Killing vector, we can canonically decompose the states into positive and negative energy and some of the usual complications of doing quantum field theory in curved space are absent (see the book by Birrell and Davis \cite{Birrell:1982ix}). 
We thus write
\begin{equation}
    A_{\text{in}\to \text{out}}= \bra {\text{out}} \text{T} \exp\left[i\int_{t_0}^{t_f} V_{\text{int}}\, \text{d} (\text{vol})\right] \ket{\text{in}}\,.
\label{eq:bulk-amplitude}
\end{equation}
The volume integral volume extends from an initial time slice (a global Cauchy slice extending all the way to the boundary) at $t_0$ to a final time slice at $t_f$. $V_{\text{int}}$ is the time dependent potential in the interaction picture (where we usually derive Feynman diagrams from). Amplitudes \eqref{eq:bulk-amplitude} are computed in perturbation theory in $V_{\text{int}}$. 

Basically, we are stating that our local Lorentz frame wave functions are global objects and we can set up perturbation theory globally in AdS.

For the flat space limit we want to show that the physics is dominated by times very close to $t\simeq 0$ and the region near $\rho\simeq 0$, but our goal is more general, so that we can also in principle talk about corrections away from the strict flat space limit.

As an example, consider a $\lambda\, (\phi(x))^4$ bulk interaction where $\phi(x)$ is an scalar field of dimension $\Delta$. $2\to 2$ scattering processes involve two incoming and two outgoing particles. To compute \eqref{eq:bulk-amplitude}, we would write the $(\phi(x))^4$ interaction in terms of raising/lowering operators for the field $\phi$. Each incoming particle would be replaced by the wave function $\psi_{\Delta, \kappa_i}(y)$, whereas each outgoing particle would be replaced by the complex conjugate wave functions $\psi^*_{\Delta, \kappa_i}(y)$, including their specific time dependence. Then we would integrate over the volume. At first order, this would look  as follows
\begin{equation}
    A_{\text{in}\to\, \text{out}}= i\lambda\, \int_{t_0}^{t_f} \psi_{\Delta, \kappa_1}(y)\psi_{\Delta, \kappa_2}(y)\psi^*_{\Delta, \kappa_3}(y)\psi^*_{\Delta, \kappa_4}(y)\,\text{d} (\text{vol}),
\end{equation}
where $\lambda$ is the perturbative coupling constant associated to the potential. 
In essence, we are doing Feynman diagrams in the bulk with wave functions for external particles at finite time. Within the LSZ formalism, one usually works with Green's functions and amputates the external legs replacing them by wave functions. Since the current wave functions are already known, there is no need to write the LSZ integrals defining them (see for example \cite{Srednicki:2007qs}, chapter 5 and \cite{Schwartz:2014sze}, chapter 6).

Since our wave functions exist in ${\BR^{2,d}}$, we can use the full set of embedding coordinates to define the integral over spacetime. We just need to add the hyperboloid constraint as a delta function. That is, for the volume measure we use the induced volume in the embedding coordinates
\begin{equation}
    \int d^{d+2} y \delta((y^{-1})^2+(y^{0})^2
- \sum (y^i)^2-1) \equiv \int d^{d+1} y^{\mu} \frac{1}{2 \sqrt{1+y^\mu y_\mu }} 
\end{equation}
where in the second integral we have resolved the delta function by calculating $y^{-1}$ near the origin, essentially assuming that it is positive. In the flat space limit, the denominator just gives $2$. These factors of $2$ can be absorbed in the definition of the perturbative coupling constant $\lambda$, or one can compensate by an extra factor of $2$ in the measure.
One just has to be careful to have a consistent rule for them.

Our next step is to understand how big a volume we should be using, considering the precise wave functions we have discussed so far. Our goal is to show that the volume is a very small region in AdS,  but a large one in the rescaled flat space limit coordinates. For that estimation, we use the fact that  the wave function is concentrated on a gaussian support of a region of order $1/\sqrt{\Delta}\simeq 1/\sqrt{\Lambda}$ and that we have a high dimension state so that the rapidity should be of order one. Once a large time of order $1/\sqrt{\Delta}$ has passed (either to the future or the past from the origin), the integrand becomes Gaussian suppressed for generic momenta (that is when the particles are moving apart with respect to each other at a velocity comparable to the speed of light). 
From then on, one is in the Gaussian tails of the distribution and integrating further in time does not contribute significantly to the amplitude. Essentially, the wave packets can be considered well separated after a small time, before the distance where curvature matters, so the region where the amplitude integral arises from is a very small region around the point $p$, even if we can only measure much later.

Now, when we attempt to take the flat space limit, we notice that the region of order $1/\sqrt{\Delta}$ in principle covers the whole of the $x^\mu$ local flat space range, since after all we have that $y= x/\Lambda$ (the $x$ has been rescaled this way to be of order one when we take the flat space limit). A distance of size $1/\sqrt{\Delta}$ in $y$, and with $\Delta \simeq O(\Lambda)$, is a distance of order $\sqrt{\Lambda}$ which is still becoming very large in the $x$ coordinates.
The amplitude becomes approximately
\begin{equation}
    A_{\text{in}\to\, \text{out}}\to i \frac{\lambda}{\Lambda^{d+1}} \int d^{d+1}x\,  \psi_{\Delta, \kappa_1}(y)\psi_{\Delta, \kappa_2}(y)\psi^*_{\Delta, \kappa_3}(y)\psi^*_{\Delta, \kappa_4}(y) ,
\end{equation}
where the range in the $x$ can be taken to infinity.
In the limit all the $\psi_\Delta$ go to plane waves and we get that
\begin{equation}
    A_{\text{in}\to\, \text{out} }\to i \frac{\lambda}{\Lambda^{d+1}}\,\delta^{d+1}(p) S_{\text{in}\to\, \text{ou}t},
\end{equation}
becomes well approximated by a delta function of momentum conservation times the flat space S-matrix. This is multiplied by a factor that goes to zero, namely  $\Lambda^{-d-1}$,  at the same time that the delta function goes to infinity. These corrections give the total amplitude as a number.
What is clear is that the limit will land exactly on the natural notion of the S-matrix. The AdS gaussians are such that they lead to well separated in and out states in the S-matrix formulation of flat space. The wave function for in and out states become the correct flat space in and out wave functions that one would use in the LSZ formalism.  The effective range of time has become infinitely long in the local flat coordinates $x$.
Similarly, spatially separated regions much farther than $O(1/\sqrt {\Delta})$ cannot contribute. The extra contribution of the spatial slices of AdS are making it so that the contribution is not only for short times near the origin, but also for short distances.
Hence the interaction integral region comes from a very small diamond near $p$, but it is written as an integral over a volume between two global Cauchy slices.

This will occur also to higher orders in perturbation theory. So long as the corresponding Feynman propagators are sufficiently suppressed at long distances, the amplitudes can be just as well computed in global  AdS as in flat space. This will be true Feynman diagram by diagram \footnote{We should note that there are various claims in the literature about the flat limit matching Feynman diagram by Feynman diagram, starting from Euclidean correlators and performing analytic continuations, for example \cite{Fitzpatrick:2011hu,Marotta:2024sce}. In our opinion, a common feature of these works is that the real time aspect that only a small region of space and time contributes is obscured.}. 

To translate between flat space and AdS we eventually need to take this limit procedure fully into account, by not being exactly at $\Lambda=\infty$, but very near that limit. The full prescription of how to do that is beyond the scope of the present paper and is being considered by the authors. Here we are just pointing out that the result from the global AdS computation is very close to the correct notion of the flat space S-matrix.

We now want to argue that the domain of integration that is optimal to study amplitudes is exactly determined by the condition 
$y^{-1} >0 $. This has to do with state preparation and measurement in the AdS spacetime itself, rather than just the flat space limit.
Since $y^{-1}= \cosh(\rho) \cos(t)$, the condition $y^{-1}>0$ translates to $\cos(t)>0$, or equivalently, 
$-\pi/2<t<\pi/2$. We want to argue that the simplest Lorentzian computation is to integrate over half of the hyperboloid for all amplitudes: this is the integral between the two times slices we just described. This region also happens to be invariant under Lorentz boosts, so the particular choice of time slicing of choosing $y^0$ versus another boosted time makes no difference. The region is determined by when the lightcone emanating from  $p$ intersects the boundary. This is independent of the precise choice of coordinates at $p$ so the amplitudes are covariant under boosts.

Consider a timelike geodesic passing through the origin $Y=(1, 0, \vec 0)$. The time $t=\pm \pi/2$ is the time required for such a geodesic to reach the aphelion (furthest point away from $\rho=0$). It is also the exact proper time of the corresponding timelike geodesic. Because $\Delta$ is large, we can use geometric optics to understand the physics of the particles away from the interaction region. These geodesics all have the same period $T= 2\pi$. The time $\pi$ is the light-crossing time of AdS, and half of that is the time to reach a turning point. 

The turning points are important because they are naturally associated with Euclidean preparation 
of the corresponding one particle state.
The idea is very simple. At the turning point all spatial momenta vanish, so it is easy to set up an analytic continuation $i\vec p_{E}= \vec p_{L}$ that keeps $p$ real both for the Euclidean and the Lorentzian theory, like is usually done in a Hartle-Hawking wave function (see also the appendix in \cite{Berenstein:2020vlp} to understand how to deal with angular momentum).

\begin{figure}[ht]
\includegraphics[width=5 cm]{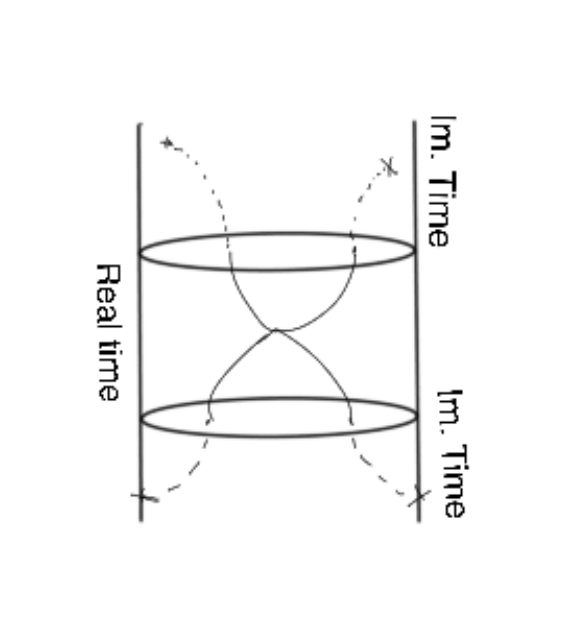}\caption{Schematic representation of a scattering event. Particles are prepared in Euclidean time. Evolution then becomes that of real time (from $t=-\frac{\pi}{2}$ to $t=\frac{\pi}{2}$) and becomes Euclidean again for readout.} \label{fig:scattering}
\end{figure}

When we reach the turning point, we can now follow the Euclidean geodesic emanating from the turning point towards the boundary (both in the future and in the past). 
We insert boundary operators ${\cal O}_\Delta$ associated to the one particle state in the bulk at the location where the spacelike geodesics end on the boundary. This will count as state preparation/readout in the CFT. The geometry can be easily visualized in figure \ref{fig:scattering}.

As was computed in \cite{Berenstein:2019tcs}, we find that the Euclidean time in the boundary (towards the past) and the radius of the turning point $\rho^*$ are related by 
\begin{equation}
    \sinh(\tau)= \frac1{\sinh(\rho^*)},
\label{eq:euc_matching}    
\end{equation}
so that there is a precise relation between the geometry of $\rho^*$ and the boundary time insertion, which is independent of $\Delta$.
Given that the energy (in AdS units) of the particle at $\rho^*$ is $E= \Delta \cosh(\rho^*) = \Delta \cosh(\eta)$ where $\eta$ is the rapidity at the origin, we find that knowledge of $\rho^*$ is equivalent to the knowledge of the rapidity in the flat space limit. 
Similarly, the value of the Euclidean time $\tau$ on the boundary completely determines $\rho^*$. The angles on the sphere are also the same, so the Euclidean time position plus the angle completely specifies the incoming or outgoing states.
A similar picture arises also in the work \cite{Komatsu:2020sag}
where the in/out states are built from boundary Euclidean caps on AdS and the boundary to bulk Green's functions take  a similar form to ours.

This Euclidean time propagation starts at $t=\pm \frac{\pi}{2}$, so a more precise statement is that we insert operators for initial states at (complex) Lorentzian times $t_{\text{in}, \alpha}= - \frac{\pi}{2} + i \tau_\alpha$ where $\alpha$ goes over the {\em in} state particles. Similarly,  $ t_{\text{out}, \beta}=  \frac{\pi}{2} -i \tau_\beta$ runs over the {\em out} state particles. This follows the analytic continuation convention where forward time in Lorentzian signature for positive energy states behaves as $\exp(-i \omega t_\mt{L})$ in the wave function. In Euclidean time it behaves as $\exp(-\omega \tau_\mt{E})$, so that the Euclidean and Lorentzian time are related by $i t_\mt{L}\simeq \tau_\mt{E} $. Since the insertion in the past of $\tau$ corresponds to negative $\tau$ values, we find that $t_{\text{in}, \alpha}$ has a negative real part and a positive imaginary part. The same argument follows for the outgoing states, where they will have a positive real part and a negative imaginary part.

Let us finally make sure the normalisation of states prepared in the boundary and in AdS are compatible. First, all the AdS states have the same norm because we just boosted the correctly normalized primary states in the bulk. The same must be true for the CFT preparation of the state that is dual to these.  Consider an operator in Euclidean conformal field theory on the cylinder. The two point function is given by
\begin{equation}
    \langle \CO(+\tau, \theta) \CO(-\tau, 0) \rangle =\frac{1}
    {|\exp(2\tau)+\exp(-2\tau) - 2 \cos(\theta)|^\Delta}
\end{equation}
This should be thought of as an overlap between two states prepared at Euclidean time $-\tau$
\begin{equation}
    \braket{\CO,\tau,\theta} {\CO,\tau,0}=\langle \CO(+\tau, \theta) \CO(-\tau, 0) \rangle
\end{equation}
evaluated in the $\tau=0$ slice.

The norm of the state is evaluated when $\theta=0$, so 
\begin{equation}
\braket{\CO, \tau,0} {\CO, \tau,0} = \frac{1}{|2 \cosh(2\tau) - 2 |^\Delta}
\end{equation}
and using $\cosh(2\tau)-1 = \cosh^2(\tau)+\sinh^2(\tau)-1 = 2 \sinh^2(\tau)$, we find that
\begin{equation}
|\ket{\CO, \tau,0}|^2 = \frac {1}{(2\sinh(\tau))^{2\Delta}}
\end{equation}
Therefore the correct prescription is to include a factor of $(2\sinh(\tau))^\Delta$ in front of the operator insertion. This makes sure that we have properly normalized the states in the boundary at $t=-\frac{\pi}{2}$  to correspond to the same normalization as states in the bulk (a similar statement is needed at $t=\frac{\pi}{2}$ for the out states). This factor $(2\sinh(\tau))^{-\Delta}$ is the exponential of the subtracted action for the spacelike geodesic that prepares the state \cite{Berenstein:2019tcs}. It is essentially $\exp(-S)$. It can be thought of as the propagator evaluated from the boundary to the turning point.

One can also ask what is special about $t=\pm \frac{\pi}{2}$ from the point of view of the variable that enters the wave functions in \eqref{eq:xi}. Since $y^{-1}\to 0$ at $t=\pm \frac{\pi}{2}$, the expression
\begin{equation}
 \xi= y^{-1}+i( \cosh(\eta)y^0+\sinh(\eta) \hat n \cdot \vec y ) 
\end{equation}
has a phase that is independent of the position in the  $t=\pm \frac{\pi}{2}$ Cauchy slice. This is the wave function notion of the particle having zero momentum (the function is essentially real).

We can also compare the boundary wave function associated to the insertion of the operator at our choice of position 
$\CO(-\tau, 0)\ket 0$ and angle.
We find that 
\begin{equation}
    \braket{\CO, 0,\theta} {\CO, \tau,0} =\frac{1}{|2\cosh(\tau) - 2 \cos(\theta)|^{\Delta}}
\end{equation}
We need to compare this to the extrapolate dictionary of our wave function (evaluated at $y^{-1}\to 0$). We find that up to a constant phase
\begin{eqnarray}
     \lim_{\rho\to \infty} \exp(\Delta \rho) \psi_{\Delta, \kappa}(t=-\frac{\pi}{2}) &=&  \lim_{\rho\to \infty} 
    \frac{\exp(\rho \Delta) } {( \cosh(\eta)y^0+\sinh(\eta) \hat n \cdot \vec y )^\Delta} \nonumber \\
    &\to&  \frac{2^\Delta}{
    (\cosh(\eta)- \sinh(\eta) \cos(\theta))^\Delta
    }\nonumber\\
    &=& \frac{2^\Delta \sinh(\eta)^{-\Delta}}{\left(\frac{\cosh(\eta)}{\sinh(\eta)}-\cos(\theta)\right)^\Delta}.
\end{eqnarray}
Here we choose $\hat n$ in the direction where the wave leaves the origin, which is the opposite from where it emanates, so we need to add a minus sign.
Notice the similarity of the angular dependence of  both prescriptions. The boundary value of the bulk wave function has a factor of $\sinh(\eta)^{-\Delta} \equiv \sinh(\rho^*)^{-\Delta} \equiv \sinh(\tau)^\Delta$, which is exactly proportional to the normalization factor we alluded to before to normalize the state correctly from the bulk perspective.
We also need that
$\frac{\cosh(\eta)}{\sinh(\eta)} = \cosh(\tau) $ in order for both to have the same functional form.
Since $\cosh(\eta) = \sqrt{1 +\sinh^2(\eta)}$, we find that there is a trigonometric identity that verifies what we need
\begin{equation}
\frac{\cosh(\eta)}{\sinh(\eta)} = (\sqrt{ 1 +\sinh(\tau)^{-2}})\sinh(\tau) = \cosh(\tau)
\end{equation}
In essence, we find that the bulk wave function and the operator insertion at a single point give the same answer and that we do not require a complicated boundary integral  prescription to build the states (we do not need to use an HKLL construction). The reason in the end is group theoretical. The rotated (boosted) operators have an associated boosted  special conformal transformation that  annihilates them. That is, $\tilde \FK \psi_{\Delta,\kappa}=0$.
The existence of such a $\tilde \FK$ indicates that the operator is exactly a primary operator with respect to some insertion on the boundary at a single (complex) value of time and boundary position.

\section{Massless particles and their wave functions}
\label{sec:massless}

The na\"\i ve procedure of starting with a primary state and doing finite boosts of the ground state fails for massless particles. The reason is that the ground state primary ends up having zero energy and is not a proper quantum of the massless field theory in flat space. Acting with a finite boost, the state remains at zero energy. An alternative way to say this is that a massless particle in flat space can never be at rest.

There are then two routes to take. We can start with flat space plane waves and figure out how to prepare the states, or start with primaries that are very easy to prepare and boost them enough to get a finite energy state in the flat space limit. Then we can take a double scaling limit and see what those wave functions look like globally.

The first one is associated to the hard preparation of scattering states \cite{Giddings:1999jq}. In principle, constructions like HKLL \cite{Hamilton:2006az} can provide the local fields, and these can be convoluted with the desired wave function to prepare the states. This is the approach of Hijano \cite{Hijano:2019qmi}.  In a sense, we would be doing the reverse of \eqref{eq:prep_states}.

We will take the second route, as we think it is more important to prioritize the easy preparation of states in the dual field theory. Because of that procedure, we start with global solutions of the field equations, whereas with plane waves we would have to work very hard to get a global solution of the field equations away from the region of interest. These will not lead exactly to the right hand side of \eqref{eq:prep_states}, but a more general class of $\ket {in}$ and corresponding $\ket{out}$ states from which the S-matrix can in principle be extracted. So long as we can write the final states easily in terms of momentum states, we can reconstruct the amplitude we want to compute from the S-matrix, or alternatively, compute the S-matrix.

Consider an AdS primary state with fixed $\Delta$ which we assume to be of order one. Since the latter does not scale with $\Lambda$, it will give rise to a massless particle in the flat space limit. Boosting the AdS particle with rapidity $\eta$, its semi-classical energy becomes  $\tilde E\simeq \Delta \cosh\eta$. We can achieve a finite flat space energy by considering the double scaling limit
\begin{equation}
  \eta \to \infty\,, \qquad E= \Delta\, \frac{e^\eta}{\Lambda}\,\,\,\text{fixed}\,.
\label{eq:massless-scaling}
\end{equation}
This describes a massless particle with energy centered around $E$.

However, the flat space limit of these AdS boosted wave functions does not give rise to ordinary plane waves. Indeed, keeping the leading exponentially large terms in $\eta$ in \eqref{eq:xi}, or in \eqref{eq:bwf}, the resulting flat space wave form becomes
\begin{equation}
\psi(\Delta, \kappa) \to \frac 1{(1- i\Lambda^{-1} \exp(\eta) \omega_\mu x^\mu )^\Delta}\equiv (1- i v_\mu x^\mu )^{-\Delta}
\label{eq:masslesswaveform}
\end{equation}
where $\omega^\mu = \frac{1}{2} (1,\,-\hat n)$ is a null vector. Notice the final answer can be written in terms of an arbitrary future pointing null vector $v^\mu$ as we did in the right hand side of \eqref{eq:masslesswaveform}. One can check this wave form solves the flat space massless wave equation. However, it is \textit{not a state at fixed energy} \footnote{The wave functions \eqref{eq:masslesswaveform} are also not translation invariant. Thus, the action of translations generates new wave functions, where the $1$ is replaced by $1+i u$, where $u$ can be thought of as a time delay parameter.}.

The wave functions \eqref{eq:masslesswaveform} are parallel waves, that is, they have a flat wave front. If the direction vectors between two such waves point in different directions, then it is easy to show that they become orthogonal by considering an analysis as in equation \eqref{eq:inner-3}, where the inner product of the $\kappa$ can be shown to scale. When the null vectors are parallel, the states are not orthogonal.

Notice that $(1- i v_\mu x^\mu )^{-\Delta}$ is a non-singular solution of the wave equation in the limit. The wave function has singularities at complex values of $x^\mu$, but it is perfectly regular in flat space. Also, for fixed $\Delta$, remember that the wave functions have a width of order one in $AdS$. The large boost has Lorentz contracted this distance to be finite in the longitudinal direction of the wave profile
fitting nicely inside the flat space region near the point $p$. 
These can be considered
very sharp waves in the direction of $\hat n$, but they are wide functions in the transverse directions to the front of the wave.
These wave functions are more similar to the ones that arise when considering celestial amplitudes (see for example the reviews \cite{Raclariu:2021zjz,Pasterski:2021rjz}), although they are not the exact same ones. For example, in our case $\Delta$ is real and positive, whereas in the celestial wave functions one usually has a parameter $\Delta$ that is complex and the one in the denominator is missing. Notice that $\Delta$ instead of becoming a mass parameter, appears in the limit as part of the shape of the wave packets. Different values of $\Delta$ are still associated to different species of massless particles since they are considered  different species in AdS after all.

Since the standard flat space S-matrix is most naturally described in terms of plane waves, it is natural to write the wave functions \eqref{eq:masslesswaveform} in terms of integrals of the latter.
Introducing the Schwinger parameter $s$, this can be easily achieved
\begin{equation}
(1- i v_\mu x^\mu )^{-\Delta}\equiv \frac{1}{\Gamma(\Delta)} \int_0^\infty ds\, s^{\Delta-1} \exp(-s(1- i v_\mu x^\mu))
\label{eq:schwinger-massless}
\end{equation}
Notice that since $v^0>0$ is the condition for the vector to be future pointing, one gets a superposition 
of positive frequency waves only. Furthermore, the very high-energy states corresponding to $s\gg1$, become exponentially suppressed and can essentially be ignored. This suggests the relevant physical amplitudes should not suffer from kinematic UV divergences. Notice that the plane waves are built only of states with momenta $k_\mu \propto v_\mu$, so vectors that point in different directions are orthogonal by the usual $\delta$ function normalization of plane waves. 

Notice that since $v_\nu$ is null, the integrand itself is a solution of the massless Klein Gordon equation. These look like a Mellin integral, so we will call them such. The expressions then have a resemblance to the Mellin amplitude formalism of Penedones et al. \cite{Penedones:2010ue,Fitzpatrick:2011ia}.

We can now try and compute amplitudes as we did in section \ref{sec:Amplitudes}. We just need to replace the integrals for time dependent perturbation theory, using the wave packets we have for in and out states.
Amplitudes involving the $\psi$ can then be rewritten as integrals over the S-matrix, with one Schwinger parameter per external particle. Indeed, if we write the in and out states in terms of waves at fixed momentum with the aid of \eqref{eq:schwinger-massless}, this statement is self-evident. We will get this way a generalized notion of Fourier transform of the S-matrix (this will be different than the celestial transform \cite{Pasterski:2016qvg}).

We want to argue that only a small region near the origin contributes. We will take that as a given in our formulae, but this is actually very subtle. The reason that it works here, unlike the case for very massive particles, is that the wave profiles decay for large spatial $x$. Two incoming wave packets that collide will both have large values of the field in some spatial hypersurface, but this does not extend in time.  Hence the time interval where the interactions happen is of small duration, even in the 
$x$ coordinates. The spatial region still goes to infinity and can extend to AdS size radii transversely. If enough of the outgoing particle wave functions intersect transversely to each other in this hyperplane, then we should be able to localize the multi-particle amplitude to a small region of flat space, finding an approximation of Landau singularities in the full AdS as in \cite{Maldacena:2015iua}
(alternatively, lightcone cuts for measurements intersecting in the bulk at the point $p$ as in \cite{Engelhardt:2016wgb}).

If we apply the same S-matrix formalism for amplitudes as in \eqref{eq:bulk-amplitude}, when replacing the wave functions by their integral transforms \eqref{eq:schwinger-massless}, one would be naturally led to perform the position $x$ integrals first. These give rise to $\delta$-function contribution
\begin{equation}
    A_{\text{massless}} \simeq \prod \int ds_i s_i^{\Delta_i-1} \exp(-s_i) \delta^{d+1}(\sum s_i v_\mu) S_{\text{massless}}(s v_{\mu,i})
\label{eq:massless-amp}
\end{equation}
depending on the Schwinger parameters. Notice the resulting massless amplitude \eqref{eq:massless-amp} is a specific Mellin integral of $\exp(-\sum s_i)\delta^{d+1}(\sum s_i v_\mu) S_{\text{massless}}(s v_{\mu,i})$ (a Mellin transform evaluated at specific values determined by the set of $\Delta$'s). For generic null momentum vectors $v_\nu$, with a small number of particles, the delta functions eliminate most, but not all of the $s$ variables. Since given a solution of the delta function support, one can obtain a different one by rescaling all the vectors by the same amount $s_i \to \gamma s_i$, we infer that after evaluating the delta function in \eqref{eq:massless-amp}, there should still be at least one Mellin integral left to perform. If one of the particles is massive, the scale can (at least in principle) be resolved by the massive particle momentum and its recoil.

We believe this produces a similar result for the flat space limit of massless particles that has been argued in these other works.  
Investigating this in detail is beyond the scope of the present paper.
Indeed, the most interesting massless particles are gravitons and we need to further describe spinning particle wave functions. 
Our purpose has been to notice that the flat space limit naturally produces non-trivial waveforms that remember $\Delta$ and that they produce Mellin-like convolutions with the S-matrix. They also have a natural action of boosts on them, and depend on a null vector that plays an analogous role to the flat space momentum $p^\mu$.

We would like to finish the discussion on massless particles by considering the AdS state preparation described in section \ref{sec:Amplitudes} to this case. Remember the energy of the AdS boosted particle is $\tilde{E} = \Delta \cosh\eta$, whereas AdS state preparation requires $\tilde{E} = \Delta \cosh\rho^*$, where $\rho^*$ is the turning point. Due to the double scaling \eqref{eq:massless-scaling}, we find $\rho^*=\eta$. Since the Euclidean insertion time $\tau$ satisfies \eqref{eq:euc_matching}
\begin{equation}
    \sinh(\tau) = \frac{1}{\sinh\rho^*} \simeq e^{-\eta} \to 0
\end{equation}
$\tau$ is infinitesimally small (of order $1/\Lambda$ in our notation). Hence, the flat space limit for massless particles on the boundary requires a double scaling limit of the insertion point of the operator, where the Euclidean time evolution goes to zero with the same speed as the flat space limit parameter $1/\Lambda$. In other words, the insertions of the corresponding operators occur almost in real time. The scattering events then look very similar to state preparation that occurs in a small band around $t=- \frac{\pi}{2}$ and readout at $t=\frac{\pi}{2}$. If we were to prepare the state \`a la HKLL, the insertion can be reduced to a very small band in space and time around the time where we match to Euclidean physics. This is the same geometric limit as in \cite{deGioia:2024yne} and corresponds to Landau singularities in the bulk \cite{Maldacena:2015iua}. Similar limits appear in constructions like the ones in \cite{Gary:2009mi}, which addresses some of the complications of thinking about the flat space S-matrix.

A more recent expression of conjectures related to the flat space limit in the language of conformal correlators can be found in \cite{vanRees:2023fcf}. These seem to depend on being able to work with complex momenta. Our description works for any $\Delta$, but the precise double scaling on the boundary that we found that produces \eqref{eq:massless-scaling} is new. Another approach starts from CFT and extracts unitarity of the S-matrix from conformal correlators \cite{Fitzpatrick:2011dm}. Again, these all are correlators at finite $\Delta$, so all particles become massless in such a limit and are subject to our geometric description in the bulk.

\section{Discussion}\label{sec:future}

This work studied some representation theory aspects of the flat space limit of AdS bulk scalar fields. Our approach is within canonical quantization rather than using path integrals and prioritises the AdS state preparation in terms of euclidean boundary operator insertions.

Using the embedding formalism, AdS primary scalar wave functions are seen to be holomorphic functions outside a disk. When taking the flat space limit in \eqref{eq:bulk-flat}, one is naturally led to consider separately the case of massive or massless particles. 

For massive particles, primary wave functions lead to plane waves at rest, while the original AdS descendants become null states. To build the entire flat space multiplet, i.e. to add moving particles, one proceeds \`a la Wigner and boosts the AdS wave functions. Since boosts are AdS isometries, these transformations can be studied in AdS, before taking the limit. As expected, such AdS boosted wave functions are \textit{not} orthogonal to the AdS primary states, but they become orthogonal
in the flat space limit. Technically, the embedding formalism allows us to compute the inner product between all relevant AdS wave functions in a manifestly $\mathbb{R}^{2,d}$ invariant way. These wave functions allow a straightforward operational AdS state preparation in terms of  operator boundary insertions with a particular complexified time on the boundary.
AdS transition amplitudes between these holographic states also give rise to Lorentz invariant  S-matrices in the flat limit. 

When following the same strategy for massless particles, we learn that to generate a \textit{finite energy} particle, one must consider a double scaling limit \eqref{eq:massless-scaling} in which the AdS scale $\Lambda \to \infty$, the boost rapidity $\eta\to \infty$ keeping $\Lambda^{-1}\,e^\eta$ \textit{fixed}. The resulting wave functions do \textit{not} lead to plane waves, they involve several energies, depend on arbitrary null vectors and remember the AdS conformal dimension $\Delta$ in their profile. These wave functions resemble similar quantities considered in euclidean AdS and look similar to some of the conformal primary wave functions considered in the celestial holography literature \cite{Raclariu:2021zjz,Pasterski:2021rjz}. Our preliminary investigation on these wave functions suggests one may be able to extract a massless S-matrix using a Schwinger parameterization that is reminiscent of the types of Mellin transforms that appeared in celestial holography. Our wave functions still allow an easy operational AdS state preparation in terms of Euclidean operator boundary insertions. In the massless limit, our state preparation recovers the na\"\i ve  geometric picture that emerged in a  HKLL approach involving real time physics in a narrow band around $t=\pm \frac{\pi}{2}$, but the full details are different. 

There are multiple directions for future research. First, we would like to understand the massless case in detail. Our massless waveform \eqref{eq:masslesswaveform} and their Schwinger parameterization \eqref{eq:schwinger-massless} must be related to the basis of massless scalar conformal primaries discussed in celestial holography (see \cite{Pasterski:2016qvg,Pasterski:2017kqt,Cotler:2023qwh}, for example)
\begin{equation}
    \varphi^\pm_\Delta(X;q) = \frac{\Gamma(\Delta)}{\pm (-i\,q\cdot X \mp \epsilon)^\Delta}\label{eq:celestialwf}
\end{equation}
and their Mellin transforms, respectively. Notice that our discussion did not account for the little group labelling of massless particle representations \`a la Wigner (our null vectors were kept generic).
Also, $\epsilon$ is a regulator in \eqref{eq:celestialwf}, whereas in our case, the equivalent quantity is finite and the values $\Delta$ can take are fixed from the AdS dual, rather than from some other physics. 
We suspect that keeping track of these matters one should be able to address both the orthogonality and completeness properties of the wave functions we constructed in \eqref{eq:masslesswaveform}  better.
We can tentatively say that our wave functions are some intermediate point between momentum eigenstates and celestial wave functions: they have features of both. They are also better behaved relative to the UV properties of the corresponding scattering states (the deep UV is exponentially suppressed). 

A second related problem is to discuss spinning particles, both massive and massless, and to take their flat space limits. This is currently being investigated. Perhaps we can better understand gravity in flat space by taking these limits carefully on AdS gravitons.

Thirdly, within a top-down approach all known examples of AdS/CFT involve a ten, or eleven, dimensional bulk. Since these involve a supergravity theory with a vanishing cosmological constant, the curvature of AdS is related to the curvature of the compact space geometrizing the R-symmetry group of the dual CFT\footnote{{For a recent discussion on this point and a different proposal, based on the carrollian structure appearing in the boundary theory, see \cite{Fontanella:2025tbs}.}}. For example, the duality between N=4 SYM and type IIB on AdS$_5\times$S$^5$, relates the $\SU(4)$ R-symmetry in the 4d CFT with the $\SO(6)$ isometry on the 5-sphere. When taking the double scaled limit described in the main text \eqref{eq:bulk-flat}, the 5-sphere becomes $\mathbb{R}^5$. After all, if interactions are required to take a very small amount of time in the AdS directions, the information of a scattering event cannot propagate causally to far distances in the 5-sphere either.

From the \`In\"on\"u-Wigner contraction perspective, there are two points to be made. First, the resulting algebra arising from the R-symmetry generates a non-compact group. Second, the total number of generators obtained from the 5-sphere symmetry contraction plus the AdS symmetry contraction does not match the generators of the ten dimensional Poincar\'e algebra.
The first remark is related to the continuum spectrum that one expects to describe the physics of flat space. Flat space should be ten dimensional, not five dimensional. Hence, from a five dimensional perspective there should be an infinite number of particles with a continuum mass. This is resolved by working correctly in 10 dimensions.
The second remark points out the existence of \textit{emergent symmetries} in the flat limit. All flat space Lorentz transformations rotating the original AdS and sphere directions into each other are emergent and responsible for the connectivity of the full ten dimensional bulk. It would be interesting to understand how the correlations and entanglement properties in the R-symmetry charged sectors of the dual theory capture the emergence of flat space. 

A fourth point to be made is that this work explored the flat space limit in the bulk from a representation theory perspective. If we were to take the same perspective on the dual boundary theory, the literature describes the corresponding \`In\"on\"u-Wigner contraction as a Carrollian limit (see \cite{Bagchi:2025vri} for a review). Relating the bulk wave functions derived in this work with their boundary counterparts may provide a constructive perspective on the meaning of the Carrollian limit and the physics of Carrollian quantum field theories. 

Finally, our entire discussion was concerned with vacuum physics of a few particles in the regime where QFT on a fixed curved spacetime makes sense. What about thermal states? Clearly, the flat Schwarzschild black hole is the flat limit of the \textit{small} AdS black hole. Just as the latter are thermodynamically unstable, black holes in flat space decay too. It would be interesting to explore whether Carrollian physics is responsible for such instabilities, extending results in 3d pure gravity \cite{Aggarwal:2025hji}.


\acknowledgments
D.B. would like to thank R. Emparan, S.Giddings, Z. Li, P. Mitra, D. Marolf,  J. Penedones, A. Puhm,  A. Raclariu, B. van Rees, S. Vandoren for various discussions related to this work. D.B. would like to thank the Institute of Physics at the University of Amsterdam.
for their hospitality while this work was being carried out. The work of D.B. was supported in
part by the Department of Energy under grant DE-SC 0011702. The work of JS was supported by the Science and Technology Facilities Council [grant number ST/X000494/1].

\bibliography{Wigner.bib}

\end{document}